# ESPSim: A JAVA Application for Calculating Electrostatic Potential Map Similarity Scores

**Gregory Martyn**[1] **& Christopher M. Frenz** [2]
[1]Department of Mathematics and Computer Science
Manhattan College
Riverdale, NY 10471
[2]Department of Computer Engineering Technology
New York City College of Technology (CUNY)
Brooklyn, NY 11201
cfrenz@citytech.cuny.edu

**Abstract -** *ESPSim is an open source JAVA program that enables the comparisons of protein electrostatic potential maps via the computation of an electrostatic similarity measure. This program has been utilized to demonstrate a high degree of electrostatic similarity among the potential maps of lysozyme proteins, suggesting that protein electrostatic states are conserved within lysozyme proteins. ESPSim is freely available under the AGPL License from* http://www.bioinformatics.org/project/?group_id=830.

## 1 Introduction

Protein electrostatics have been demonstrated to play diverse and important functional roles in proteins, including the facilitation of catalysis, ligand-binding, and protein-protein interactions, as well as protein stability. Moreover, recent findings have suggested that protein electrostatic states and their associated functionality are conserved over evolutionary time. Within the replicative proteins of HIV and HCV, a correlation between residue conservation and pKa perturbation has been demonstrated, in which the portion of residues exhibiting the greatest degree of conservation had a greater number and magnitude of perturbation than lesser conserved residues. These perturbations were exhibited in functionally important residues, and are the result of electrostatic interactions between the residue of interest and its neighboring residues [1]. Furthermore, studies on TIM-barrel proteins have demonstrated the presence of evolutionary conserved electrostatic networks, and an examination of the electrostatic surfaces of the members of 4 protein families and 1 superfamily have demonstrated the presence of conserved charge distributions on the surface of the proteins [2-3].

This growing number of studies seeking to examine the similarities present between the electrostatic states of proteins suggests the need for a methodology of comparing protein electrostatic potential maps, such as those output by the programs APBS [4], UHBD [5], Delphi [6], and MEAD [7]. Electrostatic potential map similarity measure methodologies have been developed and applied to small organic molecules for use in predicting the potential for ligand binding and in the design of transition state inhibitors [8]. These methodologies have been adapted for use with proteins and the open source JAVA program ESPSim has been written to perform the similarity calculation. JAVA was chosen as a development platform to allow for the application to run on a diversity of operating systems, and the application was written in a threaded manor to enhance performance for calculations involving large potential maps.

## 2 Methods

The first step of the developed methodology involves selecting the protein structures for which comparisons are desired and from this group selecting a reference structure. As an illustration of the programs utility, a group of lysozyme structures was selected, including the PDB files 1DKJ [9], 1HEL [10], 1HHL [11], 1JUG [12], 1LMN [13], 1LZY [14], 2EQL [15], 2IHL [16], and 1REX [17] as well as the 1JOO [18] structure of the enzyme staphylococcal nuclease to serve as a negative control. The human lysozyme structure 1REX was then selected as a reference structure. Once a reference structure is chosen, all PDB files should be aligned to this reference structure, for which an all atom alignment was performed using the Swiss PDB Viewer [19]. After the alignments are complete, coordinates for the cubic mesh that will encompass the proteins during the electrostatic calculations, along with grid spacings, should be selected so that they are large enough to encompass the largest protein in the data set. Electrostatic potential map calculations can then be performed using the decided upon spacings and coordinates. APBS was used to compute the potential maps used in the paper, but calculations could be computed using any program with related functionality.

Once the computation of electrostatic potential maps has been completed, the points within the potential map that are to be compared need to be selected. When comparing the potential maps of small organic molecules, researchers found a variety of different selection criteria useful, with some choosing to compare all points on the molecular surface, and others comparing points that were either internal or external to the molecular surface [8,20]. This study utilizes all of the points that are present on the molecular surface, as well as all external points to the surface, since research conducted by Livesay et al. [3], has suggested that electric fields surrounding the active site of the enzyme Super Oxide Dismutase (CuZnSOD) work to orient the incoming substrate of the enzyme. The inclusion of points external to the molecule would thereby allow for the inclusion of any such fields in the similarity calculation.

The similarity measure implemented in the calculation has been previously utilized to rank the electrostatic similarities of inhibitors of AMP Deaminase, Adenosine Deaminase, and AMP Nucleosidase [8] and is computed as follows:

$$S_e = \frac{\sum_{i=1}^{nA}\sum_{j=1}^{nB} E_i^A E_j^B \exp(-\alpha r_{ij}^2)}{\sqrt{\sum_{i=1}^{nA}\sum_{j=1}^{nA} E_i^A E_j^A \exp(-\alpha r_{ij}^2)}\sqrt{\sum_{i=1}^{nB}\sum_{j=1}^{nB} E_i^B E_j^B \exp(-\alpha r_{ij}^2)}}$$

where nA and nB represent the number of points in the electrostatic potential maps of proteins A and B respectively. The electrostatic potential at particular points are represented by $E_i^A$ and $E_j^B$, and $r_{ij}$ is representative of the distance present between point i and j. The value of α is used to scale the similarity measure by controlling the number of points that can interact in the potential maps being compared in a distant dependant fashion. The smaller the value of alpha the greater the number of interacting points. An alpha value of 1 was used for the reported results, but the appropriate value of alpha to use is somewhat dependant on the grid spacings used as well as the molecules being compared. In order to allow independence from any electrostatic potential mapping software packages, ESPSim program input is in a custom file format that consists of a listing of the X, Y, and Z coordinates of each point to be compared, followed by the electrostatic potential at that point.

## 3 Results and Discussion

Utilizing this procedure the similarity comparisons between the 1REX structure potential maps and the other selected protein structure potential maps yielded high electrostatic similarity scores between the lysozyme variants examined in this study (Table 1). Determination of the ClustalW [21] aligned percentage of sequence identity suggests that the electrostatic surfaces of lysozyme variants have an even greater level of conservation than the sequences that comprise the proteins. The 1REX electrostatic potential map was also compared to itself as a positive control and was correctly demonstrated to yield a similarity score of 1.00. The staphylococcus nuclease protein 1JOO potential map was used as a negative control, since it is in approximately the same size range as lysozyme but unrelated, and was only demonstrated to have a similarity score of 0.59, significantly lower than the lysozyme protein scores.

**Table 1**: Electrostatic similarity scores computed by the ESPSim program.

| Structures | Electrostatic Similarity | ClustalW Aligned % Sequence Identity |
|---|---|---|
| 1REX vs 1REX | 1.00 | 100 |
| 1REX vs 1LMN | 0.96 | 69 |
| 1REX vs 1HEL | 0.89 | 60 |
| 1REX vs 1DKJ | 0.87 | 59 |
| 1REX vs 1LZY | 0.93 | 58 |
| 1REX vs 2IHL | 0.87 | 58 |
| 1REX vs 1HHL | 0.94 | 56 |
| 1REX vs 1JUG | 0.93 | 52 |
| 1REX vs 2EQL | 0.95 | 51 |
| 1REX vs. 1JOO | 0.59 | 3 |

In all, ESPSim provides a methodology for making comparisons between protein electrostatic potential maps, and as such could provide valuable insights into studies of protein evolution, mutation induced effects on ligand binding sites, as well as the investigation of other protein properties where electrostatics plays a contributory role.